\documentstyle[psfig]{mn}

\title[Microlensing in QSO 0957+561]{Short-timescale fluctuations in the
difference light curves of QSO 0957+561A,B: microlensing or noise ?}
\author[Gil-Merino et al.]
       {R. Gil-Merino$^{1,2}$, L.J. Goicoechea$^1$, M. Serra-Ricart$^3$,  
	A. Oscoz$^3$, D. Alcalde$^3$ 
	\newauthor
	and E. Mediavilla$^3$ \\
	$^1$ Departamento de F\'{\i}sica Moderna, Universidad de Cantabria,
Avda. Los Castros s/n, E-39005 Santander, Spain \\ E-mail: 
gilmerinor@unican.es, goicol@unican.es \\
	$^2$ Lehrstuhl Astrophysik, Institut f\"{u}r Physik, Universit\"{a}t 
Potsdam, Am Neuen Palais 10, D-14469 Potsdam, Germany \\ E-mail: 
rmerino@astro.physik.uni-potsdam.de \\
        $^3$ Instituto de Astrof\'{\i}sica de Canarias,  Via L\'{a}ctea s/n, La
Laguna, E-38200  Spain \\ E-mail: mserra@iac.es, aoscoz@iac.es, 
dalcalde@iac.es, emg@iac.es}
\date{Accepted 2000 October 10}
	
\begin{document}

\maketitle

\begin{abstract}
From optical R band data of the double quasar QSO 0957+561A,B, we made 
two new difference light curves (about 330 days of overlap between the 
time-shifted light curve for the A image and the magnitude-shifted light curve 
for the B image). We observed noisy behaviours around the zero line and no 
short-timescale events (with a duration of months), where the term event refers
to a prominent feature that may be due to microlensing or another source of 
variability. Only one event lasting two weeks and rising - 33 
mmag was found. Measured constraints on the possible microlensing variability
can be used to obtain information on the granularity 
of the dark matter in the main lensing galaxy and the size of the source. In 
addition, one can also test the ability of the observational noise to 
cause the rms averages and the local features of the difference signals. We 
focused on this last issue. The combined photometries were related to a process 
consisting of an intrinsic signal plus a Gaussian observational noise. The 
intrinsic signal has been assumed to be either a smooth function (polynomial) 
or a smooth function plus a stationary noise process or a 
correlated stationary process. Using these three pictures without microlensing, 
we derived some models totally consistent with the observations. We finally 
discussed the sensitivity of our telescope (at Teide Observatory) to several 
classes of microlensing variability.
\end{abstract}


\section{Introduction}

\subsection{Microlensing caused by the Galaxy and other spirals}

Dark matter dominates the outer mass of the Milky Way. In principle, the 
population of the Galactic dark halo may include astrophysical objects as 
black holes, brown dwarfs, 
cool white dwarfs, etc., i.e., MACHOs (massive compact halo objects) with
stellar or substellar mass, as well as elementary particles (a smooth
component). Today, from microlensing surveys, we have some information about
the granular component (MACHOs). The absence of very short duration events 
implies that the dark halo cannot be dominated by planetary objects. A joint 
analysis by {\it EROS} and {\it MACHO} collaborations indicated that MACHOs in 
the mass range $10^{-7} M_{\odot} \leq M \leq 10^{-3} M_{\odot}$ make up less
than 25\% of the dark halo (Alcock et al. 1998). From a likelihood analysis,
the {\it MACHO} collaboration concluded that a population of objects of mass
$\sim$ 0.5 $M_{\odot}$ is consistent with their first two year of data. These
MACHOs with stellar mass would have an important contribution to the total mass
(Alcock et al. 1997; Gould 1997; Sutherland 1999; Mao 2000). However, very
recent results by the {\it MACHO} team, based on approximately six
years of observations, point to a relatively small mass fraction
(Alcock et al. 2000). For a typical size halo, the maximum likelihood estimates
suggest the existence of a Milky Way dark halo consisting of about 20\% MACHOs
with mass of $\sim$ 0.6 $M_{\odot}$. The {\it EROS} collaboration also agrees
with this small contribution to the dark halo by $\sim$ 0.6 $M_{\odot}$
objects (Lasserre et al. 2000). Lasserre et al. (2000) derived strong upper
limits on the abundance of MACHOs with different masses. For example, $<$ 10\%
of the dark halo resides in planetary objects. Moreover, they ruled out a
standard spherical halo in which more than 40\% of its mass is made of dark
stars with 1 $M_{\odot}$. Finally, we remark that the Milky Way dark halo
inferred from the likelihood method (best standard fits by Alcock et al. 2000)
is consistent with the {\it HST} (Hubble Space Telescope) detection of a halo
white dwarf population (Ibata et al. 1999). A population of cool white dwarfs
contributing 1/5 of the dark matter in the Milky Way could explain all new
observational results, but the hypothesis presents some difficulties (e.g., Mao
2000; Alcock et al. 2000).

The information on the nature of galaxy dark haloes is still largely based on a
local spiral galaxy (the Galaxy), and so, the study of other galaxies seems an
interesting goal.

The Einstein Cross (QSO 2237+0305) is a $z$ = 1.69 quasar lensed by a face-on 
barred Sb galaxy at $z$ = 0.0394. The time delay between the four quasar images
is expected to be less than a day (Rix et al. 1992; Wambsganss \& Paczy\'nski 
1994), and so, one can directly separate intrinsic variability from
microlensing signal. For this lens system, light rays of the 4 images pass
through the bulge of the foreground galaxy and there is a robust evidence that
microlensing events occur (e.g., Irwin et al. 1989; Wozniak et al. 2000). The
observed events may be interpreted as a phenomenon caused by the granularity of
the matter associated with the nearby spiral. Very recently, for providing an
interpretation of the {\it OGLE} Q2237+0305 microlensing light curve, Wyithe,
Turner \& Webster (2000) used the contouring technique of Lewis et al. (1993)
and Witt (1993).   

B1600+434 is another interesting gravitational mirage lensed by an 
edge-on disk galaxy. Koopmans \& de Bruyn (2000) measured the radio time delay 
between
the two images of the system and derived a radio difference light curve which 
is in disagreement with zero. They
investigated both scintillation and microlensing as possible causes of the
non-intrinsic radio variability. If microlensing is the origin of the 
"anomalous" difference light curve, then it could indicate the presence of a 
lens galaxy dark halo filled with MACHOs of mass $\geq 0.5 M_{\odot}$. 

\subsection{Microlensing in the first gravitational lens system (Q0957+561)}

A third well-known microlensed quasar is the $z$ = 1.41 double system 
Q0957+561A,B. The main lens galaxy is an elliptical galaxy (cD) at $z$ = 0.36. 
While the light associated with the image B crosses an internal region of the 
lens galaxy, the light path associated with the component A is $\approx$ 5 
arcsec away from the centre of the galaxy. The cD galaxy is close
to the centre of a galaxy cluster, and consequently, the normalized surface
mass densities $\kappa_A$ and $\kappa_B$ are the projected mass densities of
the lensing galaxy plus cluster along the lines of sight, normalized by the
critical surface mass density. Pelt et al. (1998) used the
recent values $\kappa_A$ = 0.22 and $\kappa_B$ = 1.24, which originate from an
extended galaxy halo consisting of the elliptical galaxy halo and an additional
matter related to the cluster. It is possible that a considerable
part of the extended halo mass does consist of a dark component, although an
estimate of the stellar contribution (luminous stars) to $\kappa_A$ and 
$\kappa_B$ is not so easy as in the Milky Way. For the image B, if the fraction
of mass in granular form $\kappa_{BG}$ is dominated by normal stars and dark
stars similar to the objects that have been discovered in the Galaxy (Alcock et
al. 2000), and simultaneously, the main part of the halo mass is due to a
smooth component ($\kappa_{BG} << \kappa_B$, $\kappa_{BG} <<$ 1) and the source
quasar is small, then we must expect some long-timescale microlensing event
caused by one star (luminous or dark) crossing the path of this
image. In that case of small source/one star approximation, the timescale of 
an event will be $t_o$(years) $\approx$ 17 $\sqrt{M(M_{\odot})}[600/v_t(km 
s^{-1})]$, where $v_t$ is the transverse velocity, and the magnification of the
B component has a typical duration of several years for a 0.5-1 $M_{\odot}$ star
and any reasonable choice of $v_t$. When $\kappa_{BG}$ is high ($\kappa_{BG} 
\sim$ 1) and/or the source is large, several stars at a time must be considered
and the model by Chang \& Refsdal (1984) is not suitable. The small source/one 
star model by Chang \& Refsdal (1984) was generalized in the case of a small 
source and a large optical depth (Paczy\'nski 1986) and the case of an extended 
source and an arbitrary optical depth (Kayser, Refsdal \& Stabell 1986; 
Schneider \& Weiss 1987; Wambsganss 1990). Therefore, the formalisms by Chang
\& Refsdal (1984), Paczy\'nski (1986) and Wambsganss (1990) as well as new
analytical approximations seem useful tools for a detailed analysis of the 
optical microlensing history of QSO 0957+561. A long-timescale microlensing 
signal was unambiguously observed from 1981 to 1999; see Pelt et al. (1998), 
Press \& Rybicki (1998), Serra-Ricart et al. (1999, subsequently Paper I). In 
this paper, we concentrate on the possible rapid microlensing signal. In a 
forthcoming paper, we will carry out a comparison between the annual 
differences (averages from January to December) $B - A$ and the predictions 
from different models and physical parameters. 

In the past, using a record of brightness including photometric data (in the R
band) up to 1995 and a time delay of 404 days, Schild (1996, hereafter S96) 
analyzed the possible existence of short-timescale microlensing (rapid external
variability on a timescale of months) and very rapid microlensing events (with
duration of $\leq$ 3 weeks) in the double QSO 0957+561A,B. He found numerous 
events with quarter-year and very short timescales (a few days). S96 also 
claimed that the slower component (events with a width of 90 days) 
can be interpreted as the imprint of an important population of microlenses with
planetary mass of $\sim 10^{-5} M_{\odot}$. Assuming an improved delay value of
417 days, Goicoechea et al. (1998, subsequently G98) showed a difference light
curve corresponding to the 1995/1996 seasons in Schild's dataset. G98 obtained
fluctuations which could be associated with microlensing events, in fact, our
results are in agreement with the existence of strong microlensing: the
fluctuations in the difference light curve are clearly inconsistent with zero 
and similar to the fluctuations in the quasar signal. 
New work by Schild and collaborators pointed in the same
direction: adopting a time delay of 416.3 days, Pelt et al. (1998) found that 
Schild's photometry shows evidence in favour of the presence of short-timescale
microlensing; Schild (1999) made a wavelet exploration of the QSO 0957+561
brightness record, and reported that the rapid brightness fluctuations observed
in the A and B quasar images (whose origin may be some kind of microlensing)
are not dominated by observational noise; and Colley \& Schild (1999), from a
new reduction of "old" photometric data (subtracting out the lens galaxy's light
according to the {\it HST} luminosity profile and removing cross talk light 
from 
the A and B images apertures), derived a structure function for variations
in the R-band from lags of hours to years, a time delay of 417.4 days and a 
microlensing candidate on a timescale of a day, which could imply planetary 
MACHOs in the lens galaxy halo. So, from the photometry taken at Whipple 
Observatory 1.2 m telescope by Schild, one obtains two important conclusions. 
First, there is evidence in favor of the existence of true short-timescale 
microlensing signal. Second, this rapid signal seems to support the presence of
MACHOs (in the halo of the cD galaxy) having a very small mass. However we note
that Gould \& Miralda-Escud\'e (1997) have introduced an alternative 
explanation to the possible rapid microlensing in the double QSO 0957+561A,B,
which is related to hot spots or other moving structures in the accretion disk 
in the quasar, and so, planetary objects are not involved. 

QSO 0957+561A,B was photometrically monitored at
Apache Point Observatory (Kundi\'c et al. 1995, 1997) 
in the g and r bands, during the 1995 and 1996 seasons.  Schmidt \&
Wambsganss (1998, hereafter SW98) analyzed this photometry and searched for a
microlensing signature. Considering
the photometric data in the g band and a delay of 417 days, SW98 produced a 
difference light curve covering $\approx$ 160 days and concluded that it is
consistent with zero. There is no variation in the difference light curve with 
an amplitude in excess of $\pm$ 0.05 mag and the total magnitude variation of 
a hypothetical microlensing signal is assumed to be less than 0.05 mag 
(see the dashed lines in Fig. 1 of SW98). 
They employed this last upper limit to obtain interesting
information on the parameter pair MACHO-mass/quasar-size. The lack of observed 
fluctuations rules out a population of MACHOs with $M \leq$ 10$^{-3}$ 
$M_{\odot}$ for a quasar size of 10$^{14}$ cm (25\%-100\% of the matter in
compact dark objects). However, other possible scenarios (e.g., a small source
and a halo consisting of MACHOs with $M \geq$ 10$^{-2}$ $M_{\odot}$, a source
size of 10$^{15}$ cm and a halo with compact dark objects of mass $\leq$ 
10$^{-3}$ $M_{\odot}$, etc.) cannot be ruled out from the bound on the
microlensing variability in the 160 days difference light curve. In short, SW98
have not found reliable evidence for the presence of rapid microlensing events. 

The gravitational lens system Q0957+561 was also monitored by our group with
the IAC-80 telescope at Teide Observatory, from the beginning of 1996 
February to 1998 July (see Paper I). We re-reduced our first 3 seasons 
(1996-1998) of QSO 0957+561
observations in the R band, made the difference light curves for 1996/1997
seasons and 1997/1998 seasons and studied the origin of the deviation between 
the light curves of the two images. All the results are presented in this 
article. The plan of the paper is as follows: in Sect. 2 we present the 
difference
light curves and report on new constraints on microlensing variability. In 
Sect. 3 we suggest different models that explain the difference signal. In 
Sect. 4 we discuss the sensitivity of the telescope to different microlensing 
"peaks". At the end of the paper (Sect. 5), we summarize our results.

\section{First three seasons of QSO 0957+561 observations in the R band:
difference light curves}

We have been monitoring Q0957+561 over the past 4 years (from 1996 February) 
with the 82 cm IAC-80 telescope (at Teide Observatory, Instituto de 
Astrofisica de Canarias, Spain) and have obtained a large R band dataset. 
The contribution to the solution of the old controversy 
regarding the value of the time delay ($\approx$ 400-440 days or $>$ 500 days 
?) was the first success of the monitoring program (Oscoz et al. 1996; see also
Kundi\'c et al. 1995, 1997; Oscoz et al. 1997). 

In order to give refined measurements of both time delay and optical microlensing, 
we have introduced 
some modification  with respect to the original aperture photometry (see Oscoz 
et al. 1996). Reduction of the images A and B is complicated by the presence of
cross contamination and contamination from light of the main lensing galaxy.
The two kinds of contamination depend on the seeing, and it is not clear what is
the optimal way
of obtaining the best photometric accuracy. At present, we reduce each
available night by fitting a profile to the images, which is consistent with
the point spread function of comparison stars. This new method of reduction and
the photometry from 1996 to 1998 (the first 3 seasons) are detailed in Paper I.
A table including all data is available at
\verb+http://www.iac.es/project/quasar/lens7.html+.

In the QSO 0957+561 quasar, a time delay of $\approx$ 420 days is strongly 
supported
(e.g., G98). Using our first 3 seasons of data, the time delay estimates (in
Paper I) are of 425 $\pm$ 4 days (from the $\delta^2$-test, which is based on
discrete correlation functions) and 426 $\pm$ 12 days (from dispersion
spectra). A comparison between the discrete cross-correlation function and the
discrete autocorrelation function, indicates that a time delay of $\leq$ 417
days is in disagreement with the photometry (see Fig. 16 of Paper I), while a
delay of about 425 days is favoured. Thus, we adopted a time delay of 425 days.

We concentrate now on the difference light curves. In order to estimate the 
difference light curve (DLC) for the 1996/1997 and the  1997/1998 
seasons, we used 30 observations of image A corresponding
to the 1996 season ($A96$), 28 observations of image B corresponding to the
1997 season ($B97$), 44 photometric data of image A in the 1997 season dataset
($A97$) and 84 photometric data of image B in the 1998 season dataset ($B98$).
There are about 100 days of overlap between the time-shifted (in the time
delay) light curve $A96$ and the light curve $B97$, and about 230 days of
overlap between the time delay-corrected light curves $A97$ and $B98$. 
A main problem of the IAC-80 telescope (using the available
observational time of 20-30 min/night) is related to the photometric errors.
The mean errors in the initially selected datasets are approximately 19 mmag
($A96$), 24 mmag ($B97$), 28 mmag ($A97$) and 24 mmag ($B98$). For
short-timescale microlensing studies, these errors are large and one must 
re-reduce the data (grouping them for obtaining lower errors). Because of the 
possible rapid microlensing variability on one month timescale, the 
timescale of the groups should not be too large ($\leq$ 10 days); it
should not be too small for having a sufficient number of data, and so,
relatively small errors. The re-reduced photometry consists of 12, 11, 22 and
36 "observations" in four new (and final) datasets $A96$, $B97$, $A97$ and
$B98$, respectively. These datasets are available by sending a request to 
\verb+rmerino@astro.physik.uni-potsdam.de+. 
For groups in $A96$, the timescales are less than 3 days
and the mean error is of $\approx$ 12 mmag, for groups in $B97$, the timescales
are $\leq$ 8 days and the mean uncertainty is of $\approx$ 16 mmag, for groups
in $A97$, the timescales are also $\leq$ 8 days and the mean error is lowered
to $\approx$ 20 mmag, and for groups in $B98$, the maximum timescale and the
mean uncertainty are 6 days and 16 mmag, respectively. Therefore, making groups
with a maximum timescale of $\approx$ 1 week (the mean timescale is of 
$\approx$ 2 days), the mean errors are lowered in 7-8 mmag.

\begin{figure}
\psfig{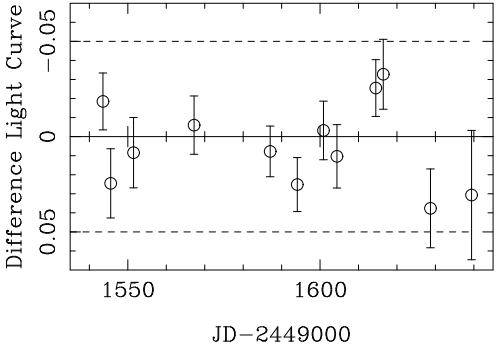}
\caption{Difference light curve for 1996/1997 seasons (in the R band). We used 
bins with semi-size of 9 days and adopted a time delay of 425 days. The times
associated with the circles are the dates in the time-delay shifted light curve
$A96$ (see main text).}
\label{Fig. 1}
\end{figure}

As we have seen in the previous paragraph, the brightness record for the A
image ($A96$ or $A97$) is measured only at a set of discrete times $t_i$ ($i$ =
1,...,N) and the light curve for the B image ($B97$ or $B98$) is also
determined at discrete times $t_j$ ($j$ = 1,...,M). Since the observational
light curves are irregularly sampled  signals, to obtain the DLC ($A96/B97$ or
$A97/B98$), we can use different methodologies, for example, the  
interpolation suggested by SW98 or the binning that appears in
G98. Here, we are interested in the DLC binned in intervals with size 2$\alpha$ 
around the dates in the light curve $A^{TS}$ (time delay-shifted light curve
$A$). In other words, each photometric measurement $A^{TS}_i$ at the date $t_i$
+ $\Delta\tau_{BA}$, where $\Delta\tau_{BA}$ is the time delay, will be 
compared to the observational data $B^{MS}_j$ = $B_j$ + $<A>$ - $<B>$ at $t_i$ 
+ $\Delta\tau_{BA}$ - $\alpha$ $\leq$ $t_j$ $\leq$ $t_i$ + $\Delta\tau_{BA}$ + 
$\alpha$ ($B^{MS}$ is the magnitude-shifted light curve $B$). The values
$B^{MS}_j$ within each bin are averaged to give $<B^{MS}_j>_i$ ($i$ = 1,...,N),
and one obtains the difference light curve (DLC)  
\begin{equation}
\delta_i = <B^{MS}_j>_i - A^{TS}_i ,
\end{equation}
being $i$ = 1,...,N. The observational process $A^{TS}$($t$) can be expanded 
as an intrinsic signal $s$($t$) plus a noise process $n_A$($t$)
related to the procedure to obtain the measurements, and a microlensing 
signal $m_A$($t$). In a similar way, $B^{MS}$($t$) = $s$($t$) + $n_B$($t$) + 
$m_B$($t$). So, the deviation $\delta_i$ must be interpreted as a combination 
of several factors, i.e.,
\begin{equation}
\delta_i = [<s_j>_i - s_i] + [<n_{Bj}>_i - n_{Ai}] + [<m_{Bj}>_i - m_{Ai}] .
\end{equation}
If $s$($t$) is a smooth function, then $s_i$ = $s$($t_i$) and $s_j$ = 
$s$($t_j$), while when $s$($t$) is a stochastic process, $s_i$ represents a
realization of the random variable $s$($t_i$) and $s_j$ denotes a realization
of the random variable $s$($t_j$). With respect to the observational noise, 
$n_{Ai}$ is a realization of the random variable $n_A$($t_i$) [similarly, 
$n_{Bj}$ is one of the possible values of $n_B$($t_j$)]. Also, in Eq. (2), 
$m_{Ai}$ = $m_A$($t_i$) and $m_{Bj}$ = $m_B$($t_j$). From Eq. (2) it is
inferred that the difference signal will be never zero, even in absence of
microlensing. There is a background dominated by observational noise, which is 
present in any realistic situation. In the case of very weak or null 
microlensing,
we expect a trend of the DLC rather consistent with zero [taking into account
the standard errors $\epsilon_1$,...,$\epsilon_N$ in the deviations estimated
from Eq. (1)]. However, in the case of strong microlensing, several absolute
deviations $|\delta_i|$ should be noticeably larger than the associated
uncertainties $\epsilon_i$. 

\begin{figure}
\psfig{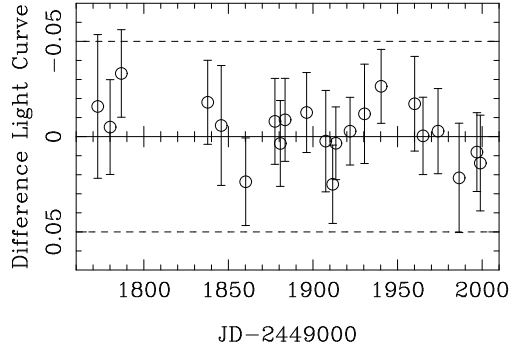}
\caption{Difference light curve for 1997/1998 seasons (in the R band). We used 
bins with semi-size of 8 days and adopted a time delay of 425 days. The
deviations [$\delta_i$; see Eq. (1)] are evaluated at discrete dates
corresponding to the time-delay shifted light curve $A97$.}
\label{Fig. 2}
\end{figure}

For the 1996/1997 seasons (from the final datasets $A96$ and $B97$), using a time
delay of 425 days and bins with semisize of $\alpha$ = 9 days, we derived the
DLC that appears in Fig. 1. Two thresholds are also illustrated: $\pm$ 0.05 mag
(discontinuous lines). In Fig. 1, there is a "peak" around day 1615: two 
contiguous points 
significantly deviated from the zero line, that verify  
$|\delta_i| > \epsilon_i$. If the whole DLC is modelled as a
single Gaussian event and the data are fitted to the model, we obtain that the
amplitude and the full-width at one-tenth maximum (FWTM) of the Gaussian law
must be $\approx$ - 33 mmag and $\approx$ 14 days, respectively (best-fit 
characterized by $\chi^2$/N $\approx$ 1). Apart from this very short duration 
event,
which is probably caused by observational noise (see next section), there is no
evidence in favor of the existence of an event on longer timescales.
We note that "event" is used in a general sense, and it may be due to
true microlensing, observational noise, a combination of both or other
mechanisms. In particular, none Schild-event (events having a width of three 
months and an amplitude 
of $\pm$ 50 mmag; see S96) is found. Although the difference signal is only 
tested during a 100 days period, to find a Schild-event belonging to a dense 
network of similar fluctuations (positive and negative), the "sampling" would 
be sufficient. In any case, from our second DLC (see here below), we must be 
able to confirm/reject the existence of a network of events with quarter-year 
timescale and an amplitude of $\pm$ 50 mmag. Finally, there are derived bounds 
on the amplitude of the microlensing fluctuations of $\pm$ 0.05 mag, which are 
similar to the bounds for 1995/1996 seasons (see Introduction and SW98). 

For the 1997/1998 seasons (from the final datasets $A97$ and $B98$), we also 
made
the corresponding DLC. In Fig. 2, the DLC and two relevant thresholds are
depicted. The difference signal is in apparent agreement with zero, i.e., Fig.
2 shows a noisy relationship $B^{MS} = A^{TS}$. We observe no Schild-events,
and therefore, {\it the total difference signal ($\approx$ 1 year of overlap
between the time-shifted light curve for the A component and the
magnitude-shifted light curve for the B component) is in clear disagreement
with the claim that 90 days and $\pm$ 50 mmag fluctuations occur almost
continuously}. One can also infer constraints on the
microlensing variability. In good agreement with the DLCs for 1995-1997
seasons, a hypothetical microlensing signal cannot reach values out of the very
conservative interval [- 0.05 mag, + 0.05 mag]. We finally remark that the
methodology introduced by SW98 (the technique of interpolation) leads to DLCs
similar to the DLCs discussed here (Figs. 1-2). 

\section{Interpretation of the difference signal}

The DLCs presented in Sect. 2 are in apparent agreement with the absence of
microlensing signal. However, to settle some doubts on the ability of the
observational noise in order to generate the observational features 
(e.g., the very rapid event in Fig. 1) and the measured variabilities (rms
averages), a more detailed analysis is needed. In this section, we are going to
test three simple pictures without microlensing. In brief, the ability of some
models for generating combined photometries and difference signals similar to
the observational ones is discussed in detail.

The observational combined photometry consists of both light curves $A^{TS}$
and $B^{MS}$. Thus, assuming that $m$($t$) = 0, the combined light curve (CLC)
must be related to a process $C$($t$) = $s$($t$) + $n$($t$). The intrinsic
signal $s$($t$) is chosen to be either a smooth function (polynomial; picture 
I) or a polynomial plus a stationary noise process (picture II) or a
correlated stationary process (picture III). In the first case (picture I), we
work with $s$($t$) = $\sum_{p=0}^{n}$ $a_pt^p$ (when the CLC is
reasonably smooth, this intrinsic signal is a suitable choice). The
polynomial law leads to $C_k$ = $\sum_{p=0}^{n}$ $a_pt_k^p$ + $n_k$ at a date
$t_k$, where $C_k$ ($k$ = 1,...,N+M) are the combined photometric data.
Considering that the process $n$($t$) is Gaussian with $<n(t)>$ = 0 and
$\sigma_n^2(t)$ = $<n^2(t)>$, and identifying the measurement errors
$\sigma_k^2$ with the noise process $\sigma_n^2(t_k)$, the probability
distribution of $n_k$ at a given time $t_k$ is $P_k(n_k,t_k)$ =
(1/$\sqrt{2\pi}\sigma_k$) exp(-$n_k^2/2\sigma_k^2$). Here, the angle brackets
denote statistical expectation values. As the random variables $n$($t_k$), $k$ 
= 1,...,N+M, are independent (the noise is uncorrelated with itself), the joint
probability distribution of the noise vector $\bf n$ = ($n_1$,...,$n_{N+M}$) is
given by
\begin{eqnarray}
P({\bf n}) = \prod_{k=1}^{N+M}P_k & \nonumber \\ 
= (2\pi)^{-\frac{N+M}{2}} 
\prod_{k=1}^{N+M}(1/\sigma_k) 
\exp \{-[C_k - \sum_{p=0}^{n}a_pt_k^p]^2/2\sigma_k^2\}  . 
\end{eqnarray}
Maximizing the likelihood function $L$ =$\ln P$ with respect to the parameters 
$a_p$, or equivalently, minimizing $\chi^2$ = $\sum_{k=1}^{N+M} [C_k - 
\sum_{p=0}^{n}a_pt_k^p]^2/\sigma_k^2$, we find a possible reconstruction of the
intrinsic signal (and thus, a model). If this procedure does not work (e.g., 
$\chi^2/dof$ is relatively large, with $dof$ = N+M-(n+1) being the number of 
degrees of freedom), we perform a fit including a stationary intrinsic noise as
an additional ingredient (picture II). This new ingredient can account for
noisy CLCs. The intrinsic noise $\eta$($t$) is taken to be Gaussian
with $<\eta(t)>$ = 0 and $\sigma_{\eta}^2(t)$ = $\sigma_{int}^2$, and moreover, 
$\eta$($t$) is uncorrelated with both $n$($t$) and with itself. Now, $C$($t$) =
$\hat{s}$($t$) + $\xi$($t$), where $\hat{s}$($t$) = $\sum_{p=0}^{n}$ $a_pt^p$ 
and
$\xi$($t$) = $n$($t$) + $\eta$($t$), and we focus on the global noise process  
$\xi$($t$). As the processes $n$($t$) and $\eta$($t$) are Gaussian and mutually
independent, their sum is again Gaussian, and the average and variance of 
$\xi$($t$) are the sums of the averages and variances of both individual noise
processes. The probability distribution of $\xi_k$ at an epoch $t_k$ can be
written as $P_k(\xi_k,t_k)$ = [1/$\sqrt{2\pi}(\sigma_k^2 +
\sigma_{int}^2)^{1/2}$] exp[-$\xi_k^2/2(\sigma_k^2 + \sigma_{int}^2)$], and the
joint probability distribution of the noise vector $\xi$ = 
($\xi_1$,...,$\xi_{N+M}$) should be $P$($\xi$) =
$\prod_{k=1}^{N+M}P_k(\xi_k,t_k)$. Finally, instead of the standard procedure
(to maximize the likelihood function), we equivalently minimize the function
\begin{equation}
\hat{\chi}^2 = \sum_{k=1}^{N+M} \{ \ln (\sigma_k^2 + \sigma_{int}^2) +
[C_k - \sum_{p=0}^{n}a_pt_k^p]^2/(\sigma_k^2 + \sigma_{int}^2) \}  .
\end{equation}
Through this method, the intrinsic signal is partially reconstructed. We find
the coefficients of the polynomial and the variance of the intrinsic noise, but
after the fit, the realizations $\eta_k$ ($k$ = 1,...,N+M) remain unknown.
However, the derived model permits us to make simulated CLCs and DLCs, since
only the knowledge of the smooth intrinsic law and the statistical properties
of the noise processes are required for this purpose.

A very different procedure was suggested by Press, Rybicki \& Hewitt (1992 a,b,
hereafter PRH92). They assumed the intrinsic signal as a correlated stationary
process. For this case III, it is possible a reconstruction of the realizations
of $s$($t$), provided that the correlation properties are known. PRH92
considered that the observational noise $n$($t$) is uncorrelated with $s$($t$)
(and with itself), and therefore, only the autocorrelation function $K_s(\tau)$
= $<\tilde{s}(t)\tilde{s}(t + \tau)>$ is needed, being $\tilde{s}(t)$ = $s(t) -
<s>$. The autocorrelation function of the intrinsic signal is not known a
priori and must be estimated through the CLC. We can relate the autocorrelation
properties to the first-order structure function $D_s^{(1)}(\tau)$ by 
\begin{eqnarray}
D_s^{(1)}(\tau) = (1/2\nu) \sum_{l,m} (s_m - s_l)^2 & \nonumber \\
\approx \frac{1}{2}<[\tilde{s}(t + \tau) - \tilde{s}(t)]^2> =
K_s(0) - K_s(\tau) ,
\end{eqnarray}
where the sum only includes the ($l$,$m$) pairs verifying that $t_m - t_l 
\approx \tau$ (the number of such pairs is $\nu$). From the CLC, one infers
(e.g., Haarsma et al. 1997)
\begin{equation}
D_s^{(1)}(\tau) \approx (1/2\nu) \sum_{l,m} 
[(C_m - C_l)^2 - \sigma_l^2 - \sigma_m^2] ,
\end{equation}
which is an evaluation of the difference $K_s$(0) - $K_s(\tau)$. As usual we
assume a power-law form for the first-order structure function, and perform a
fit to the power law. Finally, the variance of the intrinsic process $K_s$(0)
is assumed to be the difference between the variance of the CLC and a
correction due to the observational noise. The whole technique is described in
PRH92 and other more recent papers (e.g., Haarsma et al. 1997).

\begin{figure}
\psfig{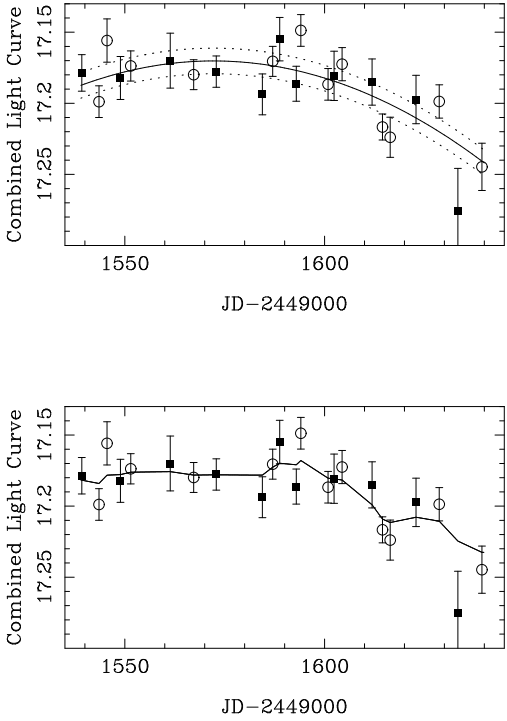}
\caption{The combined photometry of QSO 0957+561A,B for the 1996/1997 seasons 
in the R band (at Teide Observatory). The open circles trace the time-shifted 
(+ 425 days) light curve $A96$ and the filled squares trace the 
magnitude-shifted (+ 0.0658 mag) light curve $B97$. The lines are related to
two reconstructions of the intrinsic signal: considering an intrinsic signal of
the kind polynomial plus stationary noise (top panel) and the optimal
reconstruction following the PRH92 method (bottom panel).}
\label{Fig. 3}
\end{figure}

\begin{figure}
\psfig{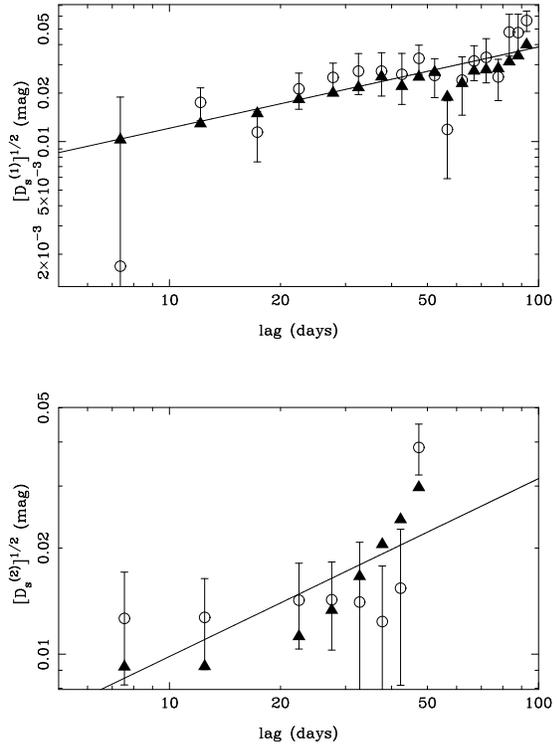}
\caption{The first-order and second-order structure functions (1996/1997 
seasons in the R band). The open circles are the values inferred from the
observational data and the filled triangles are the predictions from the
reconstruction of the kind polynomial + stationary noise. The observational 
first-order structure function was fitted to a power-law $E\tau^{\varepsilon}$ 
(solid line in the top panel). Assuming this fit as an estimation of the 
autocorrelation properties of a hypothetical correlated stationary process 
($K_s$(0) - $K_s(\tau)$), the predicted second-order structure function is 
illustrated by a solid line in the bottom panel.}
\label{Fig. 4}
\end{figure}

\subsection{The 1996/1997 seasons}

For the 1996/1997 seasons, we first have done the corresponding CLC. In a second
step, using the picture I (see above), we attempted to
fit the combined photometry. A quadratic law ($n$ = 2) gives $\chi^2/dof$ =
1.65 (best fit), whereas $\chi^2/dof$($n$ = 1) = 2.52, $\chi^2/dof$($n$ = 3) = 
1.74 and $\chi^2/dof$($n$ = 4) = 1.83. Thus the modelling of the CLC has proven
to be some difficult. Fortunately, the inclusion of an intrinsic noise
(picture II) with
moderate variance does not fail to generate an acceptable fit. When the
intrinsic signal is the previous best quadratic fit to which an intrinsic noise
with $\sigma_{int}$ = 9 mmag is added, we obtain $\chi^2/dof$ = 1.15 
($\chi^2$/N+M = 0.95). The quality of the fit has changed significantly with
the addition of the new noise, whose variance ($\sigma_{int}$ = 9 mmag) is less
than the mean variance of the observational noise (= 12-16 mmag). In Fig. 3
(top panel) the
CLC and the reconstruction are presented. The open circles represent the
time-shifted light curve $A96$, while the filled squares are the
magnitude-shifted light curve $B97$. The best polynomial ($n$ = 2) is traced by
means of a solid line, and the two lines with points are drawn at $\pm$ 9 mmag
(the best value of $\sigma_{int}$) from the polynomial. Apart from the CLC, we
checked the observational structure functions $D_s^{(1)}$ [see Eq. (6)] and 
$D_s^{(2)}$ as well as the predictions (with respect to the structure
functions) from our first succesful reconstruction. The observational
second-order structure function is computed in the following way (see
Simonetti, Cordes \& Heeschen 1985; we take a normalization factor equal to
1/6):
\begin{equation}
D_s^{(2)}(\tau) \approx (1/6\mu) \sum_{l,m,n} 
[(C_n - 2C_m + C_l)^2 - \sigma_l^2 - 4\sigma_m^2 - \sigma_n^2] ,
\end{equation}
where $\mu$ is the number of ($l$,$m$,$n$) valid triads so that $t_m - t_l 
\approx \tau$ and $t_n - t_l \approx 2\tau$. Also, the predicted structure
functions are
\begin{eqnarray}
D_s^{(1)}(\tau) \approx (1/2\nu) \sum_{l,m} 
[\hat{s}(t_m) - \hat{s}(t_l)]^2 + \sigma_{int}^2 , \nonumber \\
D_s^{(2)}(\tau) \approx (1/6\mu) \sum_{l,m,n} 
[\hat{s}(t_n) - 2\hat{s}(t_m) + \hat{s}(t_l)]^2 + \sigma_{int}^2 , 
\end{eqnarray}
being $\hat{s}$($t$) the fitted quadratic law. Fig. 4 shows the good agreement
between the observational values (open circles) and the predicted trends
(filled triangles). This result confirms that the reconstruction is reliable.
The meaning of the two straight lines in Fig. 4 will be explained here below.

\begin{figure}
\psfig{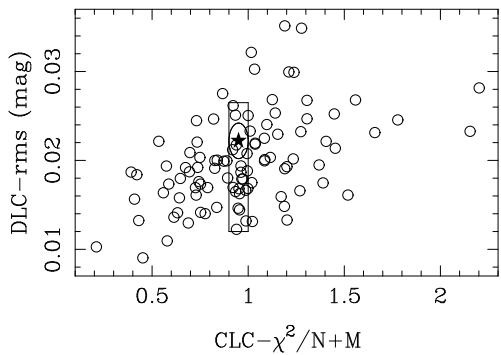}
\caption{Global properties of the measured photometry for 1996/1997 seasons
(filled star) and 100 simulated photometries (open circles). The numerical
simulations arise from M1, which is a model with three ingredients: polynomial
law + intrinsic noise + observational noise.}
\label{Fig. 5}
\end{figure}

\begin{figure}
\psfig{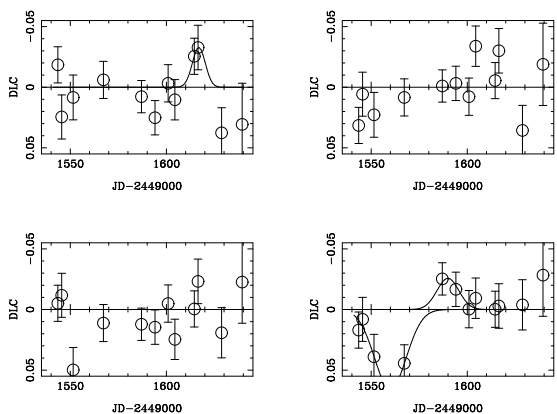}
\caption{The true DLC for 1996/1997 seasons (left-hand top panel) together with
3 simulated DLCs (via M1). The solid lines are fits to Gaussian events. A
curious result observed in the simulated DLCs is the existence of events, which
could be naively interpreted as microlensing fluctuations.}
\label{Fig. 6}
\end{figure}

\begin{figure}
\psfig{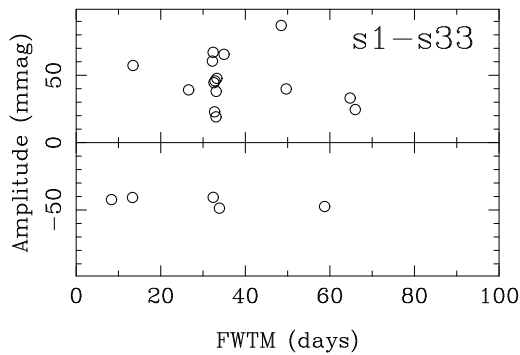}
\caption{Gaussian events (they are classified according to their amplitude and
FWTM) found in the first 33 simulations via M1. The number of features as well
as the amplitudes and time-scales are relatively surprising.}
\label{Fig. 7}
\end{figure}

Our interest in this paper is less directly in the details of a given
reconstruction of the underlying intrinsic signal than it is in analyzing
simulated photometries consistent with the reconstruction and with the 
same sampling (dates) and errors as the measured data. The first model (M1)
comprises the best quadratic fit in the absence of intrinsic noise (a smooth
component) and a Gaussian noise process characterized by a known variance at
discrete times $t_k$: $\sigma_k^2$ + $\sigma_{int}^2$. From M1 we derived 100
simulated CLCs and the corresponding DLCs. We remark that, in each simulation
(CLC), N simulated data points represent a synthetic light curve $A^{TS}$, while 
the other M data are simulated measurements of the magnitude-shifted light curve
$B$. Fig. 5 shows the relationship between the values of $\chi^2$/N+M ($\chi^2$
= $\sum_{k=1}^{N+M} [C_k - \hat{s}(t_k)]^2/[\sigma_k^2 + \sigma_{int}^2]$) and
the rms averages of the DLCs (rms = $[\frac{1}{N} \sum_{i=1}^{N}
\delta_i^2]^{1/2}$). The 100 open circles are associated with the simulated
photometries and the filled star is related to the measured photometry. The
true (measured) photometry appears as a typical result of the model. One sees
in the figure a broad range for CLC-$\chi^2$/N+M (0.2-2.2) and DLC-rms (8-36
mmag), and the true values of CLC-$\chi^2$/N+M = 0.95 and DLC-rms = 22 mmag are
well placed close to the centre of the open circle distribution. Thus, the 
measured combined photometry seems a natural consequence of M1, which is a 
model without very rapid and rapid microlensing. However, due to the event 
found in Fig. 1 (around day 1615) and other local features less prominent than
the event, we would be to doubt this conclusion and to study details in the 
synthetic DLCs. In Fig. 5, to provide some guidance, the open circles 
corresponding to 
simulated datasets with CLC-$\chi^2$/N+M similar to the measured value have 
been enclosed in a rectangular box. Also, we have drawn an elliptical surface 
centred on the filled star, which includes (totally or partially) three open 
circles associated with the synthetic photometries analogous (global properties
of both the CLC and the DLC) to the true brightness record. As we must put into
perspective the very rapid event and other local properties discovered in the 
true DLC for 1996/1997 seasons, this DLC and its features were compared with 
the three DLCs that arise from the simulations. In Fig. 6 we present the 
comparison. All events (each event 
includes a set of two or more consecutive deviations which have equal sign and 
are not consistent with zero) has been fitted to a Gaussian law and marked in 
the figure. The measured DLC (left-hand top panel and Fig. 1) is not different 
to the other three. In fact, in our 100$^{th}$ simulation (s100; right-hand 
bottom panel), two events appear. The positive event is more prominent 
than the negative event, and this last one is similar to the measured one. With
respect to the regions without events, the true variability cannot be
distinguished from the simulated ones. To throw more light upon the problem, 
we searched for Gaussian events in 1/3 of all simulations (s1-s33), as a sample
of the whole set of simulations because the computation turned out to be very
time-consuming. The results
are plotted in Fig. 7: amplitude of each event (mmag) vs. FWTM (days). There 
are a lot of events with amplitude in the interval [- 50 mmag, + 90 mmag] and 
duration $<$ 70 days. In particular, the probability of observing a negative 
event is of 15\% and the probability of observing one or more events is of 
about 50\%. So, it must be concluded that {\it the noisy (around zero) 
difference light curve based on observations is totally consistent with M1, and
the deviations from the zero line can be caused by the combined effect of the 
processes $n$($t$) (main contribution) and $\eta$($t$)}. 

\begin{figure}
\psfig{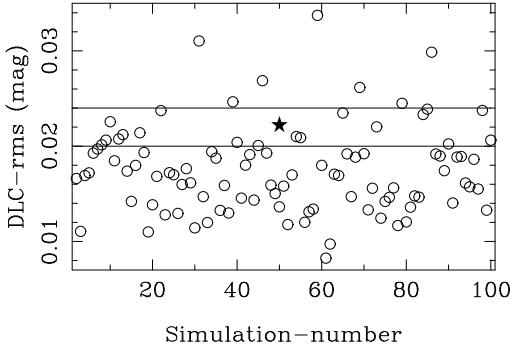}
\caption{Global properties of the true DLC for 1996/1997 seasons (filled star) 
and 100 simulated DLCs (open circles). The numerical simulations were made 
through a model including the optimal reconstruction of a correlated stationary 
process and a Gaussian observational noise process whose variance at the dates
of the real data is known.}
\label{Fig. 8}
\end{figure}

\begin{figure}
\psfig{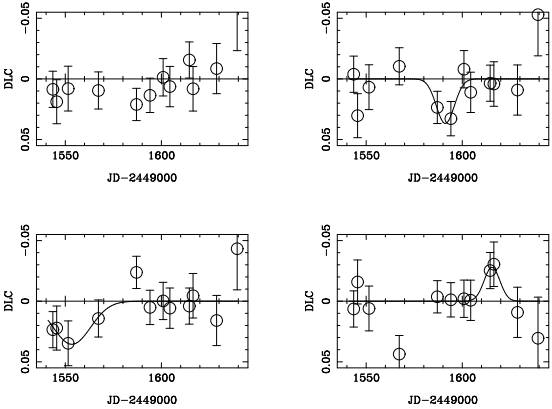}
\caption{Four simulated DLCs (via M2). For comparison with the true event in
Fig. 1 (see also Fig. 6), the Gaussian events have been clearly marked on the
panels.}
\label{Fig. 9}
\end{figure}

\begin{figure}
\psfig{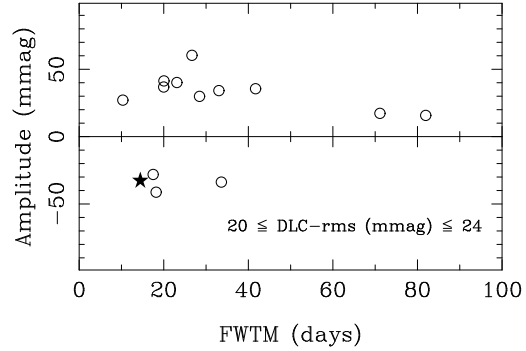}
\caption{Gaussian events found in the numerical simulations (via M2) with 20
$\leq$ rms (mmag) $\leq$ 24 (open circles). Events very similar to the real
event (filled star) are produced in two simulations.}
\label{Fig. 10}
\end{figure}

We also propose a reconstruction of the underlying intrinsic signal as
realizations of a correlated stationary process (picture III). The observational
first-order structure function can be fitted to a power-law 
$E\tau^{\varepsilon}$ (see Fig. 4, solid line in the top panel). If one 
considers this fit as an estimate of the difference $K_s$(0) - $K_s(\tau)$
for a correlated stationary process, then it is straightforward to obtain the
predicted second-order structure function (see Fig. 4, solid line in the bottom
panel: the prediction is irrelevant to reconstruct the intrinsic signal, but it
is necessary for testing the consistency of the starting point $K_s$(0) - 
$K_s(\tau)$ = $E\tau^{\varepsilon}$) and to apply the reconstruction formalism 
by PRH92. Therefore, we are able to find the realizations of the intrinsic 
process at the observational times $t_k$ ($k$ = 1,...,N+M)
as well as in the gaps between the observations. The PRH92 technique leads to
an acceptable fit with $\chi^2/dof$ = 1.18 ($dof$ = N+M-1), and our second
successful reconstruction is showed in Fig. 3 (bottom panel). The 
knowledge of both the optimal
reconstruction and the properties of the Gaussian observational noise process
at discrete times $t_k$ ($k$ = 1,...,N+M), permits us to make 100 new
simulations. In Fig. 8 details of the rms averages of the DLCs are provided
(open circles). The observational DLC has a rms average (filled star) similar
to the rms average of about 1/5 (20\%) of the simulated DLCs. Furthermore, 
four simulated DLCs with rms in the interval [20 mmag, 24 mmag] (in Fig.
8, this range of variability is labeled with two horizontal lines) appear in 
Fig. 9. From the new model (M2), DLCs with no
events (as in the analysis presented above, the Gaussian events are related to
"peaks", or in other words, we only made events around consecutive multiple
deviations with equal sign and well separate from zero) and DLCs that
incorporate more or less prominent features are derived. We note that one DLC
(right-hand bottom panel) has an event almost identical to the true one in Fig.
1 and Fig. 6. Fig. 10 shows the
properties of all Gaussian events in the simulated DLCs with rms in the 
vicinity of the observational rms (open circles). The measured event is also 
depicted (filled star), and we can see two simulated events analogous to it. We
finally conclude that {\it the observational DLC is in clear agreement with M2,
and so, microlensing would be not advocated. In this framework (M2), the 
observational noise process is a sufficient mechanism for originating the 
measured deviations}.

\begin{figure}
\psfig{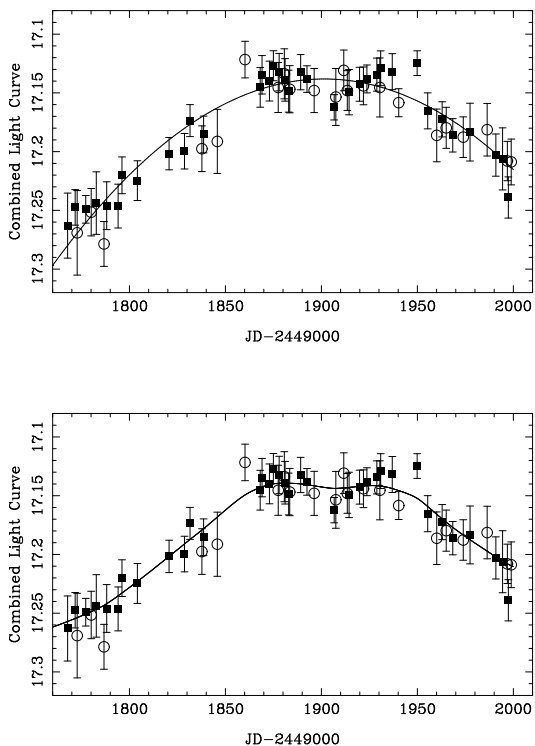}
\caption{The combined photometry for 1997/1998 seasons in the R band (at
Teide Observatory). The open circles trace the time-shifted (+ 425 days) light 
curve $A97$ and the filled squares trace the magnitude-shifted (+ 0.0603 mag) 
light curve $B98$. The solid lines represent two reconstructions of the 
intrinsic signal: the best quadratic fit (top panel) and the optimal
reconstruction following the PRH92 method (bottom panel).}
\label{Fig. 11}
\end{figure}

\begin{figure}
\psfig{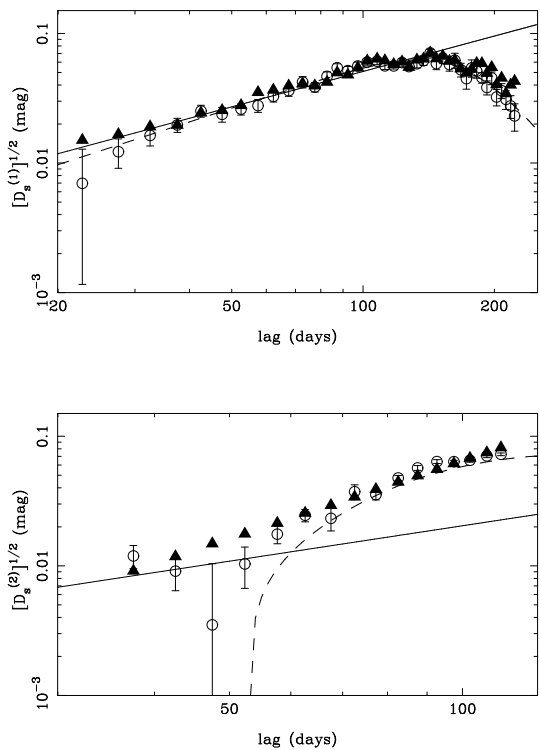}
\caption{The first-order and second-order structure functions (1997/1998 
seasons in the R band). The open circles are the values inferred from the
observational data and the filled triangles are the predictions from the
reconstruction of the kind polynomial. The observational first-order structure 
function was fitted to different laws, and two "reasonable" fits are drawn in 
the top panel (dashed and solid lines). If the fits are interpreted as the
difference $K_s$(0) - $K_s(\tau)$ for a correlated stationary process, the 
corresponding predicted second-order structure functions are illustrated by two
lines in the bottom panel.}
\label{Fig. 12}
\end{figure}

\subsection{The 1997/1998 seasons}

The combined photometry for 1997/1998 seasons and the reconstruction based on a
polynomial fit are showed in Fig. 11 (top panel). The open circles 
represent the 
time-shifted light curve $A97$ and the filled squares are the magnitude-shifted
brightness record $B98$. There is no need for the presence of an intrinsic 
noise,
and a simple quadratic law works well, leading to $\chi^2/dof$ = 0.85 (best
fit). In Fig. 11 (top panel), the solid line traces the reconstruction of
the intrinsic
signal. Besides the comparison between the measured CLC and the fitted
polynomial, we tested the predicted structure functions. In Fig. 12 we present
the observational $D_s^{(1)}$ and $D_s^{(2)}$ [open circles; see Eqs. (6-7)]
and the predictions from the best quadratic fit (filled triangles; see Eqs. (8)
with $\sigma_{int}$ = 0). The laws traced by the dashed and solid lines in this
figure will be discussed below. It is evident that the behaviours deduced from
observations and the predicted trends agree very well, and this result
indicates that the reconstruction is robust. 

\begin{figure}
\psfig{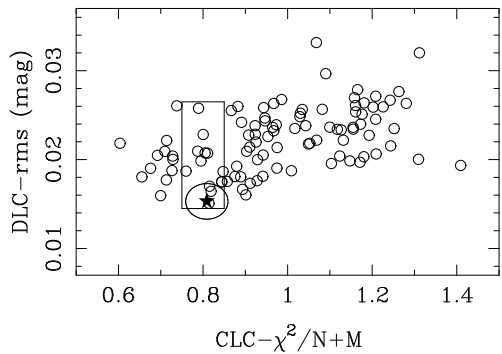}
\caption{Global properties of the true photometry for 1997/1998 seasons (filled
star) and 100 simulated photometries (open circles). The numerical simulations 
are based on a model of the kind polynomial plus observational noise.}
\label{Fig. 13}
\end{figure}
  
\begin{figure}
\psfig{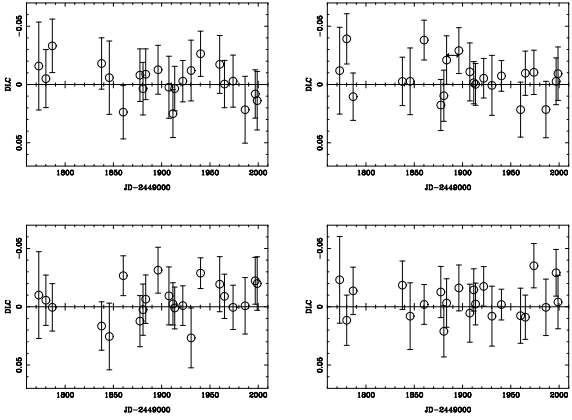}
\caption{The true DLC for 1997/1998 seasons (left-hand top panel) together with
3 simulated DLCs (via M3). An only "peak" is marked by a double arrow (see 
right-hand top panel).}
\label{Fig. 14}
\end{figure}

The first model for 1997/1998 seasons (M3) consists of the best quadratic fit
together with a Gaussian observational noise process (whose variance is 
$\sigma_k^2$ at discrete times $t_k$, $k$ = 1,...,N+M, being $\sigma_k$ the 
measurement errors at the dates of observation $t_k$). Using M3 we performed
100 simulated CLCs (and consequently, 100 simulated DLCs). The global
properties of the simulated photometries (open circles) and the true dataset
(filled star) are depicted in Fig. 13. If we concentrate on the simulations
with $\chi^2$/N+M similar to the measured value (rectangular box), the true DLC
has a rms relatively small (of about 15 mmag), but consistent with the rms 
distribution associated with the simulated DLCs. We remark that 3 simulations 
(open circles in the elliptical surface around the filled star) are analogous 
to the real brightness record, and in Fig. 14, their DLCs can be compared with 
the true DLC. The measured difference signal (left-hand top panel and Fig. 2) 
is a quasi-featureless trend and similar to the other synthetic DLCs. There are
no significant events in these four DLCs with small global variability. We  
conclude that {\it a model with no microlensing (M3) has the ability of 
generating ligth curves like the real data for 1997/1998 seasons}. Henceforth, 
we are going to treat the "peaks" as top-hat fluctuations, 
i.e., given a "peak" including deviations $\delta_{P1}$,...,$\delta_{PP}$ at 
times $t_{P1}$,...,$t_{PP}$, the amplitude and duration of the associated 
top-hat profile will be evaluated as the average of the individual deviations 
and the difference $t_{PP} - t_{P1}$, respectively. In Fig. 14, a "peak" 
(defined by two contiguous negative deviations, which are inconsistent with 
zero) appears in the DLC from the 7$^{th}$ simulation (s7; right-hand top 
panel). The "peak" is marked by a double arrow that represents the amplitude 
and duration of the associated top-hat profile. 
    
The inspection of the observational first-order structure function (see Fig.
12) suggests that the underlaying law could be intricate. To find the 
autocorrelation properties of a possible and plausible correlated stationary 
process causing the main part of the observed signal (picture III), this 
observational structure function was firstly fitted to a non-standard law 
$D_s^{(1)}(\tau)$ = $E\tau^{\varepsilon}/[1 + (\tau/T)^{\lambda}]^2$. As showed
in Fig. 12 (dashed line in the top panel), the fit is excellent. However, when 
we attempt to reproduce the observational second-order structure function, an 
inconsistent prediction is derived (dashed line in the bottom panel). The 
prediction fails at $\tau <$ 70 days. Other functions led to fits more or less 
successful, and finally we adopted the point of view by PRH92. In Fig. 12 (top 
panel) one sees a power-law behaviour up to $\tau$ = 140 days. The drop at the 
largest lags is due to the coincidence of values in the starting and ending 
parts of the measured CLC. Therefore, we assume that the observational 
first-order structure function is a reliable estimator of $K_s$(0) - 
$K_s(\tau)$ at $\tau \leq$ 140 days, whereas it is a biased estimator at $\tau 
>$ 140 days. The power-law fit to the data at lags $\tau \leq$ 140 days gives
the autocorrelation properties for the correlated stationary process, shown as
a solid line in the Fig. 12 (top panel). The predicted second-order structure 
function (Fig. 12, solid line in the bottom panel) is consistent with the
observational one up to a lag of 70 days, and it deviates from the
observational trend at $\tau >$ 70 days. However, since the observational 
second-order structure function at lag $\tau$ is associated with the 
autocorrelation at lag $2\tau$, the observational $D_s^{(2)}$($\tau >$ 70 days)
will be related to the autocorrelation at $\tau >$ 140 days, which is poorly 
traced from observations. Thus the deviation at largest lags is reasonable and 
the global prediction should be considered as a consistent result.

\begin{figure}
\psfig{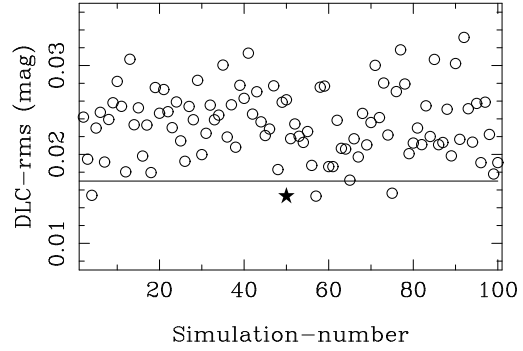}
\caption{Global properties of the true DLC for 1997/1998 seasons (filled star) 
and 100 simulated DLCs (open circles). The numerical simulations were made from
M4 (see main text).}
\label{Fig. 15}
\end{figure}

\begin{figure}
\psfig{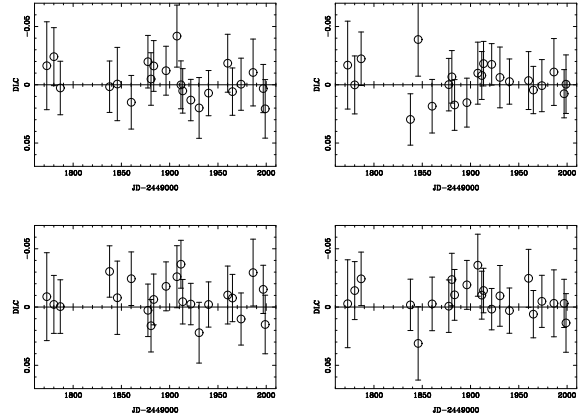}
\caption{Four simulated DLCs via M4. No events are found (for a comparison, see
Fig. 2).}
\label{Fig. 16}
\end{figure}

Once the relationship between the structure function and the autocorrelation
has been established, we can directly obtain both an optimal reconstruction of 
the realizations of the intrinsic signal and a new model (M4). The relatively 
smooth reconstruction is showed in Fig. 11 (bottom panel; the $\chi^2/dof$ 
value is of 0.86),
and the associated model leads to 100 simulations, whose global properties
(rms averages of the DLCs) are presented in Fig. 15 (open circles). In Fig. 15,
a filled star represents the true rms average, which is consistent (although
marginally) with the rms distribution from simulations. Finally, four simulated
DLCs with rms $\leq$ 17 mmag (in Fig. 15, the upper limit of 17 mmag is marked 
with one horizontal line) have been selected for a more detailed inspection. We
found noisy behaviours around zero and no events in these synthetic DLCs, i.e.,
the results agree with the analysis of the real difference signal for 1997/1998
seasons. The 4 quasi-featureless simulated DLCs appear in Fig. 16. We again 
showed that {\it microlensing is not necessary. The real combined photometry 
and 
difference signal can be due to a set of realizations of two very different 
processes: a correlated stationary process (intrinsic) and a Gaussian noise 
(observational)}.

\section{The ability of the IAC-80 telescope to detect microlensing "peaks"}

\begin{figure}
\psfig{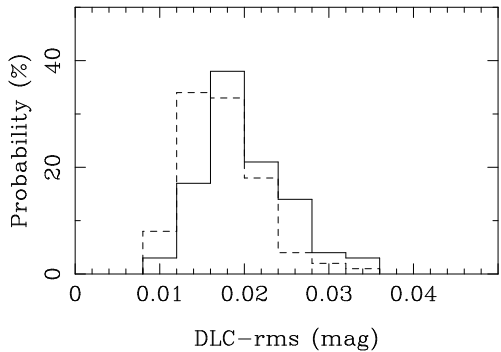}
\caption{Probability distributions of the rms averages of the synthetic DLCs.
The numerical simulations were made from M1 (solid line) and M2 (dashed line).}
\label{Fig. 17}
\end{figure}

\begin{figure}
\psfig{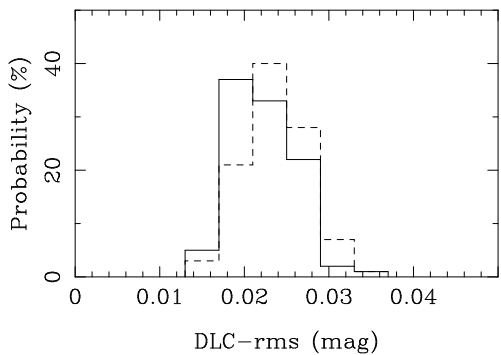}
\caption{Probability distributions of the rms averages of the synthetic DLCs.
The numerical simulations are based on M3 (solid line) and M4 (dashed line).}
\label{Fig. 18}
\end{figure}

The sensitivity of our telescope to microlensing variability in a given
observational DLC is an important issue which merits more attention. To 
explain the observations for 1996/1997 seasons and 1997/1998 seasons, we 
proposed (in Sect. 3) four models based on pictures including only an intrinsic
signal and observational noise. The simulations arising from these models 
(100 simulated difference light curves per model) are a useful tool to study 
the statistical properties of the expected difference signal in the absence of 
microlensing, and so, to test the resolution of the IAC-80 telescope for 
microlensing variability. In Fig. 17 we present the probability distributions of
the rms values (DLCs) derived from M1 (solid line) and M2 (dashed line). A 
value of about 20 mmag has a relatively high probability of 20-40\%, while a rms
exceeding 36 mmag is inconsistent with both models, as can be seen in Fig. 17.
Fig. 18 also shows the probability of observing (in the absence of 
microlensing) different rms values: via M3 (solid line) and via M4 (dashed 
line). The rms averages in the interval 19-27 mmag are highly probable
(20-40\%), but a global variability characterized by either rms $\leq$ 12 mmag 
or rms $\geq$ 38 mmag can be excluded. As a general conclusion, the rms of the
difference signal induced by noise does not exceed a threshold of 37 mmag. 
Therefore, {\it the rms values of future observational DLCs can be used to
discriminate between the presence of the expected background (global 
variability with rms $<$ 40 mmag) and the probable existence of true
microlensing signal (rms $\geq$ 40 mmag)}. 

\begin{figure}
\psfig{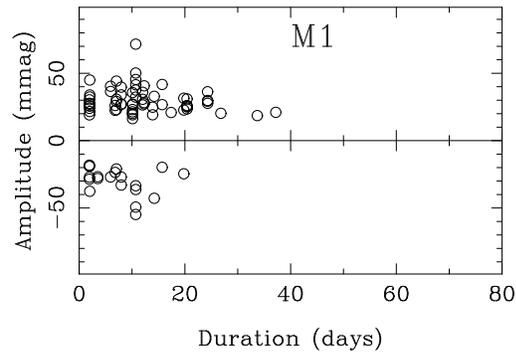}
\caption{Top-hat fluctuations found in the numerical simulations based on M1. 
We show 84 features that appear in 100 simulated DLCs.}
\label{Fig. 19}
\end{figure}

\begin{figure}
\psfig{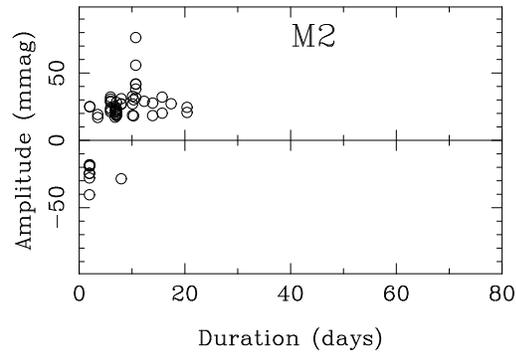}
\caption{Top-hat fluctuations in 100 simulated (via M2) DLCs. They were found 
55 "peaks".}
\label{Fig. 20}
\end{figure}

\begin{figure}
\psfig{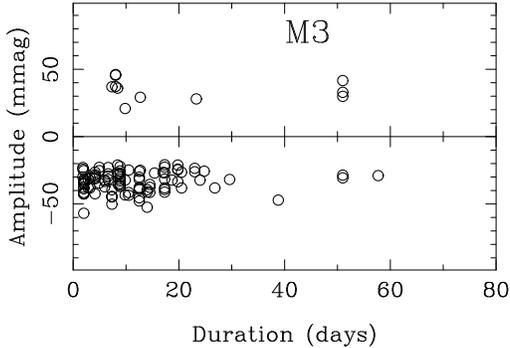}
\caption{Top-hat fluctuations from M3. We note the existence of noise "peaks" 
with a duration longer than 40 days. All these features are however associated
with an unfortunate small gap in our photometry.}
\label{Fig. 21}
\end{figure}

\begin{figure}
\psfig{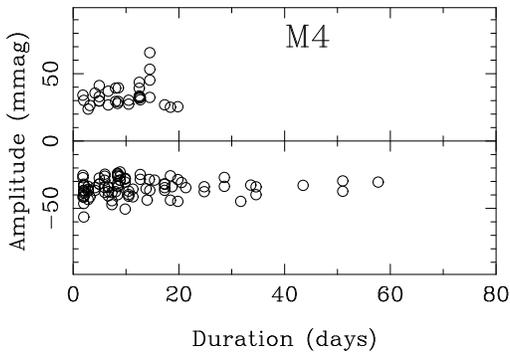}
\caption{"Peaks" from M4.}
\label{Fig. 22}
\end{figure}

The previous discussion on the global variability is interesting, but it is not 
the main goal of this section. Our main goal lies in discussing the sensitivity
of the telescope (taking into account typical sampling, photometric errors,
re-reduction and making of bins) to several classes of microlensing "peaks" 
(the cores of the microlensing events). We have seen, in Sect. 3, a figure
that shows the properties of the Gaussian events (amplitude and FWTM) found in 
a subset of simulations from M1 (see Fig. 7). Fig. 7 can be compared to the 
distribution of top-hat fluctuations found in all DLCs generated with M1. In 
Fig. 19 the distribution of the top-hat fluctuations (basically the properties 
of the "peaks" associated with them) appears, and a direct comparison between 
Fig. 7 and Fig. 19 indicates the logical fact that Gaussian fits lead to longer
durations than top-hat estimates. In the case of Gaussian fits, events with a 
duration (FWTM) of 1-2 months are abundant and only features with a timescale 
$>$ 70 days are ruled out. However, the "peaks" (from M1) with a timescale
of about one month are scarce. To discuss the power of resolution of the
telescope for local microlensing variability we chose the top-hat fluctuations
("peaks") instead of the events. The properties of an event (around a "peak")
depend on the assumed profile (e.g., Gaussian, Lorentzian, etc.) and the global
behaviour of the DLC, whereas the top-hat shape directly traces the "peaks",
avoiding to make assumptions on their wings and the use of the rest of the
corresponding DLCs. In a few words, the top-hat structures are more local
and free from assumptions than the events.

The "peaks" from M2 (Fig. 20) are not so numerous as the top-hat fluctuations 
inferred from the first model (M1). Moreover, the new cloud of points (open 
circles) is more concentrated towards shorter durations. In fact, all "peaks" 
have a timescale of $\leq$ 20 days. When one takes M3 (Fig. 21) and M4 (Fig. 
22) the situation is also somewhat different. The probability of observing a 
40-60 days top-hat fluctuation is now of about 5\%, although most features are
due to a small gap of about 50 days around day 1815 (see Paper I and Fig. 2).
Finally, Figs. 19-22 inform on the true ability of the IAC-80 telescope to 
detect microlensing fluctuations in an observational DLC free from gaps: {\it a 
"peak" with a timescale $>$ 40 days should be interpreted as a feature 
related to microlensing or other mechanisms different to the observational 
noise, while as mainly caused by the poor resolution at the expected amplitudes
within the interval [- 50 mmag, + 50 mmag], the $\leq$ 20 days microlensing 
"peaks" cannot be resolved. Even in the unlikely case of very short-timescale
microlensing signal with high amplitude, due to the smoothing by both the
re-reduction and the binning as well as the current uncertainty of one 
week in the true time delay, it would be not possible to reliably reconstruct 
the microlensing "peaks"}. 

\section{Conclusions}

Several $\sim$ 1 metre class telescopes around the world are at present involved in
different optical monitoring programs of quasars with the goal to detect microlensing. 
There are at least two "modest" telescopes searching for microlensing signal related to a far
elliptical galaxy (which is responsible, in part, for the gravitational mirage
Q0957+561). The data taken at Whipple Observatory 1.2 m telescope and at Teide
Observatory IAC-80 telescope together with the photometry from a 3.5 m
telescope (at Apache Point Observatory) represent a great effort in order to
obtain an accurate time delay in Q0957+561, follow the long-timescale
microlensing event in that system and find some evidence in favour of very
rapid and rapid microlensing (Kundi\'c et al. 1995, 1997; Oscoz et al. 1996,
1997; Pijpers 1997; Schild \& Thomson 1997; Paper I; S96; Pelt et al. 1998;
SW98; G98).

With respect to the very rapid (events with a timescale $\leq$ 3 weeks) and
rapid (events with a duration of 1-4 months) microlensing, the previous results
(before this article) are puzzling. The combined photometries (CLCs) from data
taken at Whipple Observatory only can be well explained in the context of a
picture including intrinsic variability, observational noise and microlensing
variability on different timescales: from days to months (e.g., S96). The
long-timescale microlensing does not play any role in a CLC. In
particular, S96 reported on the existence of a network of rapid events with a 
few months timescale and an amplitude of about $\pm$ 50 mmag (these features 
found by Schild are called Schild-events). However, SW98 concluded that a 
picture with intrinsic signal and observational noise (without any need to 
introduce very rapid and rapid microlensing) is consistent with the 
observations at Apache Point Observatory. SW98 really show a difference
light curve in global agreement with the zero line, but some doubt remains  
on the ability of the observational noise for producing the negative and 
positive measured events around "peaks" (a "peak" is constituted by a set of 
two or more consecutive deviations which have equal sign and are not consistent
with zero). In any case, SW98 observed no Schild-events.

In this paper, motivated by the mentioned intriguing results on microlensing
variability, we analyzed the data from our initial monitoring program with the 
IAC-80 telescope (see Paper I). We focused on the possible presence of rapid 
microlensing events in the light curves of QSO 0957+561 and the sensitivity of 
the telescope (using typical observational and analysis procedures) to 
microlensing "peaks". Our conclusions are:

\begin{enumerate}
\item Using photometric data (in the R band) for the 1996-1998 seasons, we made 
two difference light curves (DLCs). The total difference signal, which is based
on $\sim$ 1 year of overlap between the time-shifted light curve for the A 
component and the magnitude-shifted light curve for the B component, is in 
apparent agreement with the absence of microlensing signal. We can reject the 
existence (in our DLCs) of events with quarter-year timescale and an amplitude 
of $\pm$ 50 mmag, and therefore, Schild-events cannot occur almost 
continuously. On the contrary, they must be either rare phenomena (originated 
by microlensing or another physical process) or, because two observatories 
(Apache Point Observatory and Teide Observatory) found no Schild-events, untrue
fluctuations associated with the observational procedure and/or the reduction 
of data at Whipple Observatory.
\item From a very conservative point of view, in our data, the amplitude of 
any hypothetical
microlensing signal should be in the interval [- 50 mmag, + 50 mmag]. The rms 
averages of the DLCs (global variability) are of about 22 mmag (1996/1997 
seasons) and 15 mmag (1997/1998 seasons), and reasonable constraints on the 
possible microlensing variability lead to interesting information on the 
granularity of the dark matter in the main lensing galaxy (a cD elliptical 
galaxy) and the size of the source (QSO). Thus the set of bounds derived from
1995-1998 seasons (SW98 and this work) rules out an important population of 
MACHOs with substellar mass for a small quasar size (Schmidt 1999). 
\item In order to settle any doubt on the ability of the observational noise
for generating the global (rms averages) and local (events and other less 
prominent features) properties of the DLCs, we have also carried out several 
experiments as "Devil's advocates". The measured variability (the rms value, a 
very rapid event and some minor deviations) in the DLC for 1996/1997 seasons 
can be caused, in a natural way, by the observational noise process. In the 
absence of microlensing signal, we proposed two different models (M1 and M2; 
see subsection 3.1) whose associated photometries (simulations) are consistent 
with the observations. In addition, the DLC for 1997/1998 seasons is a 
quasi-featureless trend with relatively small rms average. To explain the 
variability in our second observational DLC, we again showed that microlensing 
is not necessary. Two new models (M3 and M4; see subsection 3.2) only including 
the reconstruction of the intrinsic signal (assumed as a polynomial or a 
correlated stationary process) and a Gaussian observational noise process, led 
to simulated DLCs in agreement with the measured behaviour.
\item We finally show that from a typical monitoring with our telescope 
(observing times, method of analysis, etc.) is not possible the resolution 
of $\leq$ 20 days microlensing
"peaks". The confusion with noise does not permit the separation between true 
microlensing features and "peaks" due to the observational noise. However, all
hypothetical "peaks" with a timescale $>$ 40 days must be interpreted as 
phenomena which are not associated with the observational noise (e.g., 
microlensing fluctuations). At intermediate timescales (of about one month) the
situation is somewhat intricate. Given a measured DLC, the probability of 
observing one noise "peak" (with a duration of about 30 days) is less than
10\%. Therefore, if we search for microlensing signal and find an 
"intermediate peak",
the relative probabilities that the fluctuation is a noise feature or a 
microlensing "peak" are $<$ 1:10.  
\end{enumerate}

\section*{Acknowledgments}

We are especially grateful to Joachim Wambsganss and Robert Schmidt for helpful
discussions and comments on a first version of the paper. This work was 
supported by the P6/88 project of the Instituto de Astrofisica de Canarias 
(IAC), Universidad de Cantabria funds, and DGESIC (Spain) grant PB97-0220-C02.

\bsp

\end{document}